\documentclass[conference]{IEEEtran}
\usepackage{cite}
\usepackage{amsmath,amssymb,amsfonts}
\usepackage{algorithmic}
\usepackage{graphicx}
\usepackage{textcomp}
\usepackage{xcolor}
\usepackage{booktabs}
\usepackage{multirow}
\usepackage{hyperref}
\hypersetup{colorlinks,allcolors=black}

\def\BibTeX{{\rm B\kern-.05em{\sc i\kern-.025em b}\kern-.08em
    T\kern-.1667em\lower.7ex\hbox{E}\kern-.125emX}}
\begin{document}

\title{Improving Robustness of Diffusion-Based Zero-Shot Speech Synthesis via Stable Formant Generation}

\author{
\IEEEauthorblockN{Changjin Han}
\IEEEauthorblockA{
\textit{DeepBrain AI}\\
Seoul, South Korea \\
colin@deepbrain.io}
\and
\IEEEauthorblockN{Seokgi Lee}
\IEEEauthorblockA{
\textit{DeepBrain AI}\\
Seoul, South Korea \\
fredrick@deepbrain.io}
\and
\IEEEauthorblockN{Gyuhyeon Nam}
\IEEEauthorblockA{
\textit{DeepBrain AI}\\
Seoul, South Korea \\
noah@deepbrain.io}
\and
\IEEEauthorblockN{Gyeongsu Chae}
\IEEEauthorblockA{
\textit{DeepBrain AI}\\
Seoul, South Korea \\
gc@deepbrain.io}
}

\maketitle

\begin{abstract}
Diffusion models have achieved remarkable success in text-to-speech (TTS), even in zero-shot scenarios. Recent efforts aim to address the trade-off between inference speed and sound quality, often considered the primary drawback of diffusion models. However, we find a critical mispronunciation issue is being overlooked. Our preliminary study reveals the unstable pronunciation resulting from the diffusion process. Based on this observation, we introduce StableForm-TTS, a novel zero-shot speech synthesis framework designed to produce robust pronunciation while maintaining the advantages of diffusion modeling. By pioneering the adoption of source-filter theory in diffusion TTS, we propose an elaborate architecture for stable formant generation. Experimental results on unseen speakers show that our model outperforms the state-of-the-art method in terms of pronunciation accuracy and naturalness, with comparable speaker similarity. Moreover, our model demonstrates effective scalability as both data and model sizes increase.
Audio samples are available online: \url{https://deepbrainai-research.github.io/stableformtts/}.
\end{abstract}


\section{Introduction}
In pursuit of seamless interaction between humans and computers, there has been a surge of interest in text-to-speech (TTS) synthesis. Particularly, diffusion models \cite{sohl2015deep, ho2020denoising, song2020score} have shown notable progress in conventional two-stage TTS, which converts text input into corresponding acoustic representation (e.g., mel-spectrogram) with acoustic models \cite{popov2021grad, jeong2021diff}, and then reconstructs them into waveform using  vocoders \cite{chen2020wavegrad, kong2020diffwave}. The recent progress in resolving the trade-off between inference speed and sound quality of diffusion-based TTS promotes its real-world application \cite{chen2022resgrad, huang2022prodiff, chen2023lightgrad}.

Zero-shot approaches are blossoming trends that require only a single sample of a target speaker, driven by the demand for custom voice synthesis \cite{huang2022generspeech, casanova2022yourtts, wang2023neural, jiang2023mega}. Among diffusion-based zero-shot TTS models \cite{chen2023diffusion, shen2023naturalspeech, kang2023grad}, Grad-StyleSpeech (GSS) \cite{kang2023grad} demonstrates excellent performance by enjoying the benefit of the non-autoregressive method based on the feed-forward transformer (FFT) \cite{vaswani2017attention, li2019neural}, alignment modeling \cite{ren2020fastspeech, badlani2022one}, style encoding \cite{min2021meta}, and score-based diffusion model \cite{song2020score}.

\begin{figure}[t]
    \centering
    \includegraphics[width=\linewidth]{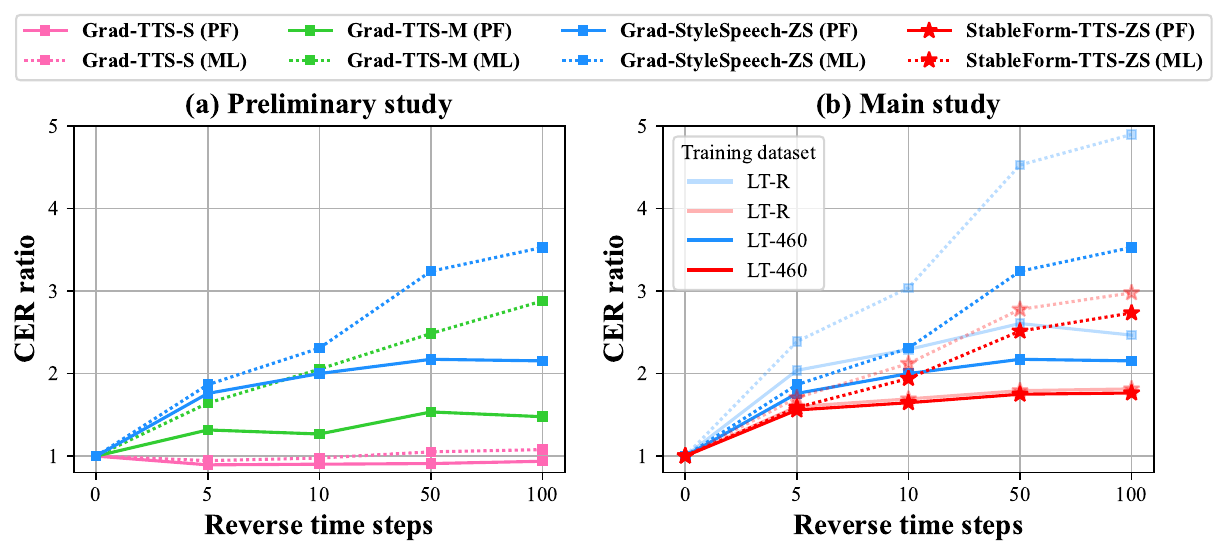}
    \caption{Line charts of CER ratio ($\frac{CER_{n}}{CER_{0}}, n\in[0,100]$) against reverse steps for each diffusion TTS model. Capital letters attached to model names stand for S: single speaker, M: multi speaker, and ZS: zero-shot, respectively. We denote solver and train dataset names in parentheses as abbreviations.}
    \label{fig:preliminary_study}
\vspace{-0.6cm}
\end{figure}

However, we empirically find that diffusion-based models suffer from a catastrophic mispronunciation problem in zero-shot scenarios, despite the robustness in a single-speaker setting \cite{popov2021grad}.
Based on our preliminary study (Fig. \ref{fig:preliminary_study}a), we reveal that there is drastic degradation in pronunciation accuracy correlated to these factors: 1) complexity of target data distribution, which is highest in zero-shot scenarios; 2) diffusion stochasticity, which accumulates with the number of reverse steps, especially in SDE solver. Additionally, we observe that the phonetic signals with weak magnitude or contrast, particularly formants, can be damaged (Fig. \ref{fig:visual_comparison}).

Meanwhile, the source-filter theory \cite{chiba1941vowel, fant1960acoustic, tokuda2021source} elucidates the mechanism of human speech production into disentangled control of a source sound and a filter. The pitch, i.e. fundamental frequency, is primarily determined by the excitation signal of the source, resulting from the vibrations of the vocal folds, while the formant frequencies are shaped by the vocal tract's filter. FastPitchFormant \cite{bak2021fastpitchformant} first adopted the source-filter theory in a neural acoustic model and successfully synthesized intelligible speech across varying sizes of pitch shift scale. 

\begin{figure*}[t]
  \centering
  \includegraphics[width=\linewidth]{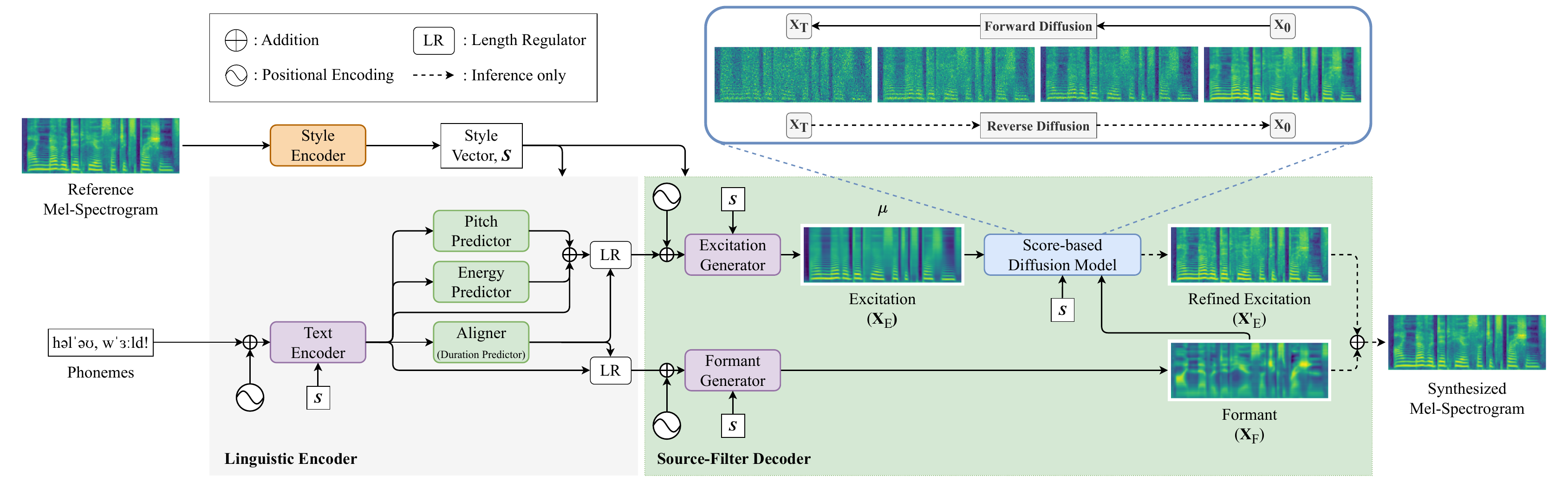}
  \caption{Overall architecture of StableForm-TTS. For brevity, the phoneme, pitch, and energy embedding layers are omitted.}
  \label{fig:stableform_tts}
\vspace{-0.6cm}
\end{figure*}

In this paper, we propose \textit{StableForm-TTS}, a diffusion-based zero-shot TTS model that focuses on pronunciation stability. To resolve the adverse effect of the diffusion model on pronunciation, we incorporate two inductive biases: 1) variance features that alleviate the over-smoothing problem of speech signals \cite{ren2022revisiting}; 2) novel architecture based on the source-filter theory. To the best of our knowledge, this paper is the first attempt to exploit the source-filter theory in diffusion TTS. 

Experimental results show that StableForm-TTS outperforms the baseline, and ablation studies verify the effectiveness of our method. Scalability tests demonstrate that our model can achieve comparable performance to state-of-the-art open-source zero-shot TTS models with significantly fewer parameters.

\section{StableForm-TTS}
\label{section:method}
In this section, we introduce StableForm-TTS tailored to generate speech with stable pronunciation. To embed appropriate inductive biases, StableForm-TTS comprises three core modules: a style encoder, a linguistic encoder, and a source-filter decoder. We illustrate the overall architecture in Fig. \ref{fig:stableform_tts}.

\subsection{Style Encoder}
To effectively convert a reference speech into a latent style vector, we utilize the mel-style encoder derived from Meta-StyleSpeech \cite{min2021meta}. The style vector encapsulates various attributes such as speaker identity, prosody, recording settings, and other pertinent features for capturing the nuanced elements of speech. We represent the style vector as $\boldsymbol{s} \in \mathbb{R}^{d_s}$ where ${d_s}$ is the hidden dimension. The reference speech matches the target speech during training but differs during inference.

\subsection{Linguistic Encoder}
The linguistic encoder generates two hidden sequences with different information from the text input. This module consists of a text encoder and a decomposed variance adaptor.

\subsubsection{Text Encoder}
The text encoder transforms the phoneme embedding sequence to the phoneme hidden sequence through the multiple FFT blocks. Following \cite{min2021meta, kang2023grad}, we adopt the style-adaptive layer normalization (SALN) receiving the style vector to compute and apply gains and biases within the FFT blocks instead of the conventional layer normalization. 

\subsubsection{Decomposed Variance Adaptor}
Similar to FastPitchFormant \cite{bak2021fastpitchformant}, we introduce a decomposed structure into the variance adaptor \cite{ren2020fastspeech, lancucki2021fastpitch}. This module has five submodules: a pitch predictor, an energy predictor, an aligner, a duration predictor, and a length regulator.

To produce two distinct representations—one enriched with prosody and the other preserving non-prosodic information—we split the variance adaptor into two pathways. Specifically, the excitation pathway adds the phoneme-level pitch and energy embedding sequences to the phoneme hidden sequence, while the formant pathway solely handles the phoneme hidden sequence.

The ground truth pitch and energy are determined by averaging the F0 values and the L2-norms of the magnitude of each frame over every phoneme, respectively. In training, we use the ground truth to train the pitch and energy predictors proposed in \cite{ren2020fastspeech, lancucki2021fastpitch}, and employ the predicted values in inference.

The aligner and the duration predictor aim to get duration information for the length regulation during training and inference, respectively. We use the aligner proposed in \cite{badlani2022one} that learns the alignment between the target mel-spectrogram and the phoneme hidden sequence using the Viterbi algorithm. Concurrently, the duration predictor is trained to predict the target duration of each phoneme extracted from the aligner. The length regulator upsamples the phoneme-level representations using the predicted durations.

\subsection{Source-Filter Decoder}
The source-filter decoder succeeds the dual pathways of the decomposed variance adaptor to transform the linguistic encoder outputs into a couple of mel-spectrograms, where one contains excitation (source) information and the other contains formant (filter) information. This module includes an excitation generator, a formant generator, and a score-based diffusion model. We only apply the diffusion model to the excitation pathway for variable prosody generation while the formant pathway is separated from the diffusion process to retain non-prosodic content. 

\subsubsection{Excitation and Formant Generators} 
Our design of excitation and formant generators (E-F generators) is motivated by \cite{bak2021fastpitchformant}. Each generator receives the asymmetric information from the excitation or formant pathways and is induced to generate corresponding representations as shown in Fig. \ref{fig:stableform_tts}. We also adopt SALN within both generators. We denote the excitation representation as $\boldsymbol{X}_E \in \mathbb{R}^{D \times T}$, and the formant representation as $\boldsymbol{X}_F \in \mathbb{R}^{D \times T}$ where $D$ is the dimension of mel-spectrograms different from \cite{bak2021fastpitchformant}, and $T$ is the number of frames.

\subsubsection{Score-based Diffusion Model}
We utilize the score-based diffusion model to enhance speech quality and diversity. The diffusion model \cite{sohl2015deep, ho2020denoising, song2020score} involves two processes: forward and reverse. The forward process gradually distorts the original data $\boldsymbol{X}_{0} \sim p(\boldsymbol{X}_{0})$ to a simple prior, typically a standard Gaussian, by adding random noise. Conversely, the reverse process learns to progressively reconstruct the data from Gaussian noise in order to form a generative model. 

Following the use of data-driven prior in TTS research \cite{popov2021grad, kang2023grad}, we define the forward and reverse diffusion processes in terms of stochastic differential equations (SDEs). The forward diffusion process can be formulated as follows:
\begin{align}
    d\boldsymbol{X}_{t} = \frac{1}{2}\beta_{t}(\boldsymbol{\mu} - \boldsymbol{X}_{t})dt + \sqrt{\beta_{t}}d\boldsymbol{W}_{t},
    \label{equation:eq1}
\end{align}
where $t \in [0, T]$ is the continuous time step, $\beta_{t}$ is a non-negative noise schedule, $\boldsymbol{\mu}$ is the data-driven prior, and $\boldsymbol{W}_{t}$ is the Brownian motion. Due to the conditional distribution $p(\boldsymbol{X}_{t}|\boldsymbol{X}_{0})$ is Gaussian, 
the forward diffusion converts any data distribution into $\mathcal{N}(\boldsymbol{\mu}, \boldsymbol{I})$ as a terminal distribution \cite{song2020score}.

The reverse diffusion process can be stated as follows:
\begin{equation}
\begin{split}
    d\boldsymbol{X}_{t} = 
    &\Bigl(\frac{1}{2}(\boldsymbol{\mu} - \boldsymbol{X}_{t}) - \nabla \log{p(\boldsymbol{X}_{t})}\Bigl)\beta_{t}dt + \sqrt{\beta_{t}}d\boldsymbol{\widetilde{W}}_{t},
\end{split}
\label{equation:eq2}
\end{equation}
where $\boldsymbol{\widetilde{W}}_{t}$ is the reverse-time Brownian motion, $\nabla \log{p(\boldsymbol{X}_{t})}$ is the gradient of the log-density of noisy data referred to as \textit{score}. 
Since (\ref{equation:eq2}) has the same marginal probability densities with following ordinary differential equation (ODE):
\begin{align}
    d\boldsymbol{X}_{t} = \frac{1}{2}\Bigl((\boldsymbol{\mu} - \boldsymbol{X}_{t}) - \nabla \log{p(\boldsymbol{X}_{t})}\Bigl)\beta_{t}dt,
    \label{equation:eq3}
\end{align}
we can use numerical SDE or ODE solver to generate samples $\boldsymbol{X}_{0}$ from noise $\boldsymbol{X}_{T}$. Note that calculating the exact score of distribution is intractable. Therefore, we train an U-Net based neural network $\boldsymbol{\varepsilon}_{\phi}(\boldsymbol{X}_{t},t,\boldsymbol{\mu},\boldsymbol{s}, \boldsymbol{X}_F)$ to estimate the score of $p(\boldsymbol{X}_{t})$ for every time step. We concatenate $\boldsymbol{s}$ and $\boldsymbol{X}_F$ channel-wise to $\boldsymbol{\mu}$ as condition, after adjusting dimension size according to $D$ with projection layers. 

Unlike previous diffusion-based TTS works \cite{popov2021grad, chen2023lightgrad, kang2023grad}, we only pass the excitation representation $\boldsymbol{X}_E$ as $\boldsymbol{\mu}$ while keeping the formant representation $\boldsymbol{X}_F$ unchanged. Thus, our StableForm-TTS synthesizes the final output $\boldsymbol{\hat{X}}$ by adding the refined excitation representation $\boldsymbol{X}'_E$ and the formant representation during inference:
\begin{align}
    \boldsymbol{\hat{X}} = \boldsymbol{X}'_E + \boldsymbol{X}_F.
    \label{equation:eq4}
\end{align}

\subsection{Training Objective}
To calculate the prediction error of the duration $\mathcal{L}_{d}$, pitch $\mathcal{L}_{p}$ and energy $\mathcal{L}_{e}$, we adopt MSE loss as in FastSpeech 2 \cite{ren2020fastspeech}.
The alignment loss $\mathcal{L}_{align}$ \cite{badlani2022one} contains CTC-based forward sum loss and KL-divergence loss between soft alignments and hard alignments to train the aligner.
As a reconstruction loss, we use $ \mathcal{L}_{prior} = {\lVert \boldsymbol{\mu} - (\boldsymbol{X} - \boldsymbol{X}_F) \rVert}^{2}_{2} $, where $\boldsymbol{X}$ is the target mel-spectrogram. We also formulate diffusion loss $\mathcal{L}_{diff}$ to optimize the score estimation network $\boldsymbol{\varepsilon}_{\phi}$ as follows: 
\begin{equation}
\begin{split}
    \mathcal{L}_{diff} = \mathbb{E}_{\boldsymbol{X}_0,t} \Bigg[ \boldsymbol{\lambda}_t\mathbb{E}_{\boldsymbol{\epsilon}_t} &{\Big\lVert  \boldsymbol{\varepsilon}_{\phi}(\boldsymbol{X}_{t},t,\boldsymbol{\mu},\boldsymbol{s},\boldsymbol{X}_F) + \frac{\boldsymbol{\epsilon}_t}{\sqrt{\boldsymbol{\lambda}_t}}\Big\rVert}^{2}_{2} \Bigg] , \\
    \boldsymbol{\epsilon}_t \sim \mathcal{N}(\boldsymbol{0}, \boldsymbol{I}), 
    \quad &\boldsymbol{\lambda}_t = \boldsymbol{I} - e^{-\int^{t}_{0} \beta_s ds}.
\end{split}
\label{equation:eq5}
\end{equation}

Consequently, the total loss for training StableForm-TTS is:
\begin{align}
    \mathcal{L}_{total} = \mathcal{L}_{d} + \mathcal{L}_{p} + \mathcal{L}_{e}+ \mathcal{L}_{align} + \mathcal{L}_{prior} + \mathcal{L}_{diff}.
    \label{equation:eq6}
\end{align}

\begin{table*}[t]
    \caption{English zero-shot TTS results on VCTK}
    \label{tab:model_comparison}
    \centering
    \resizebox{\linewidth}{!}{
        \begin{tabular}{l|l|cccccc|cc|c}
        \toprule
        \multirow{2}{*}{\textbf{Dataset}} & \multicolumn{1}{l|}{\multirow{2}{*}{\textbf{Model}}} & \multicolumn{6}{c|}{Objective Evaluation}  & \multicolumn{2}{c|}{Subjective Evaluation} & \multicolumn{1}{l}{\multirow{2}{*}{\textbf{\# Params}}}\\
        & \multicolumn{1}{l|}{}                                                          
        & \textbf{CER ($\downarrow$)} & \textbf{WER ($\downarrow$)} & \textbf{SECS ($\uparrow$)}
        & \textbf{UTMOS ($\uparrow$)} & \textbf{WVMOS ($\uparrow$)} & \textbf{MOSA-Net+ ($\uparrow$)}
        & \textbf{MOS ($\uparrow$)} & \textbf{SMOS ($\uparrow$)} & \multicolumn{1}{l}{} \\ 
        \midrule
-                                      & Ground Truth
                                       & 1.66 ± 0.14   & 3.75 ± 0.30   & 80.95 ± 0.21
                                       & 4.065 ± 0.007 & 4.402 ± 0.009 & 4.031 ± 0.008 
                                       & 3.81 ± 0.13   & 3.71 ± 0.12   & - \\
-                                      & Ground Truth (voc.)               
                                       & 1.82 ± 0.15   & 4.05 ± 0.32   & 80.41 ± 0.21
                                       & 3.845 ± 0.009 & 4.299 ± 0.010 & 4.014 ± 0.008 
                                       & 3.79 ± 0.12   & 3.63 ± 0.14   & - \\ 
        \midrule
        \multirow{2}{*}{LT-460}        & GSS
                                       & 2.73 ± 0.07   & 6.49 ± 0.14   & \textbf{79.45 ± 0.06}
                                       & 3.958 ± 0.004 & 4.443 ± 0.004 & 3.873 ± 0.004 
                                       & 3.73 ± 0.04   & 3.66 ± 0.05   & 33.98M \\
                                       & StableForm-TTS (ours) 
                                       & \textbf{1.44 ± 0.05}   & \textbf{3.64 ± 0.10}   & 78.64 ± 0.07
                                       & \textbf{4.131 ± 0.003} & \textbf{4.545 ± 0.004} & \textbf{4.072 ± 0.004}
                                       & \textbf{3.76 ± 0.04}   & \textbf{3.68 ± 0.05}   & 34.86M \\
        \midrule
        \multirow{2}{*}{LT-R}          & GSS
                                       & 5.80 ± 0.11   & 11.95 ± 0.21  & \textbf{78.45 ± 0.07}
                                       & 3.694 ± 0.004 & 4.189 ± 0.004 & 3.593 ± 0.004  
                                       & 3.67 ± 0.05   & 3.67 ± 0.05   & 33.98M \\
                                       & StableForm-TTS (ours)
                                       & \textbf{3.04 ± 0.08}   & \textbf{6.74 ± 0.15}   & 78.43 ± 0.07
                                       & \textbf{3.976 ± 0.004} & \textbf{4.360 ± 0.004} & \textbf{3.844 ± 0.004}
                                       & \textbf{3.76 ± 0.04}   & \textbf{3.70 ± 0.05}   & 34.86M \\
        \bottomrule
        \end{tabular}
    }
\vspace{-0.4cm}
\end{table*}

\section{Experiments}
\subsection{Experimental Setup}
\subsubsection{Datasets}
Except for the pre-trained models\def\thefootnote{2}\footnote{\url{https://github.com/huawei-noah/Speech-Backbones/tree/main/Grad-TTS}}, i.e. Grad-TTS-S and Grad-TTS-M used in the preliminary study, we train all the models on two datasets. One is LibriTTS \texttt{train-clean-460} (LT-460) \cite{zen2019libritts} following \cite{kang2023grad}, and the other is all subsets  of LibriTTS-R (LT-R) \cite{koizumi2023libritts}, to include phonetically noisier samples while ensuring high audio quality. For the evaluation on unseen speakers, we utilize VCTK \cite{yamagishi2019vctk} and use all of the samples from 11 speakers following previous zero-shot TTS studies \cite{casanova2021sc, casanova2022yourtts}. In the pre-processing stage, we convert the text inputs into phonemes using Phonemizer \cite{Bernard2021} with the eSpeak NG\def\thefootnote{3}\footnote{\url{https://github.com/espeak-ng/espeak-ng}} backend, and resample the audio to 22,050 Hz. Also, we extract the 80-dim mel-spectrograms with a fast Fourier transform size of 1024, a window size of 1024, and a hop size of 256. We obtain the F0 values by PRAAT toolkit \cite{boersma1993accurate}. 

\subsubsection{Implementation Details}
To maintain our model's capacity comparable to \cite{kang2023grad}, we use four FFT blocks for the text encoder and two FFT blocks for each of the excitation and the formant generators. The components in the mel-style encoder and the decomposed variance adaptor are the same as the original papers \cite{min2021meta, badlani2022one, ren2020fastspeech, lancucki2021fastpitch}. We adopt the U-Net architecture and the noise schedule introduced in Grad-TTS \cite{popov2021grad} for diffusion modeling, and the temperature $\tau$ is set to 1.5, i.e. $\boldsymbol{X}_T \sim \mathcal{N}(\boldsymbol{\mu}, \frac{\boldsymbol{I}}{1.5})$ for inference. For the reverse process, we employ the probability flow (PF) \cite{popov2021grad} as an ODE solver and the maximum likelihood (ML) \cite{popov2021diffusion, kang2023grad} as an SDE solver.
We train the models for 1 million steps on a single NVIDIA A100 GPU with a batch size of 16. We use the Adam optimizer and the learning rate schedule \cite{vaswani2017attention} with 4,000 warmup steps. The pre-trained HiFi-GAN \cite{kong2020hifi} serves as a vocoder to convert the generated mel-spectrograms into waveforms.

\subsubsection{Evaluation Metrics}
For objective evaluation, we calculate the character error rate (CER) and the word error rate (WER) to assess robustness leveraging the CTC-based conformer ASR model\def\thefootnote{4}\footnote{\url{https://catalog.ngc.nvidia.com/models}}. We adopt the speaker embedding cosine similarity (SECS) using the pre-trained speaker encoder from Resemblyzer\def\thefootnote{5}\footnote{\url{https://github.com/resemble-ai/Resemblyzer}} to assess speaker similarity. Also, we employ MOS prediction models (UTMOS \cite{saeki2022utmos}, WVMOS \cite{andreev2023hifi}, MOSA-Net+ \cite{zezario2024mosa}) to evaluate speech quality. We denote the average of the MOS predictions as MOS-pred. For subjective evaluation, we randomly select 10 speakers from the test set and conduct human ratings through Amazon Mechanical Turk\def\thefootnote{6}\footnote{\url{https://www.mturk.com/}}, involving 20 participants. These participants are instructed to rate the audio samples presented with matching texts on a 5-point Likert scale across three tasks: 1) mean opinion score (MOS) for naturalness assessment; 2) similarity MOS (SMOS) to evaluate speaker similarity; and 3) comparative MOS (CMOS) for comparing naturalness. We filtered out responses where the response time was shorter than the audio length. In addition, loudness normalization is applied to all audio samples to alleviate volume-related bias. Unless otherwise stated, we compute the metrics by averaging across two solver types (PF and ML) and four reverse steps (5, 10, 50, and 100) with 95\% confidence intervals.

\begin{figure}[t]
    \centering
    \includegraphics[width=\linewidth]{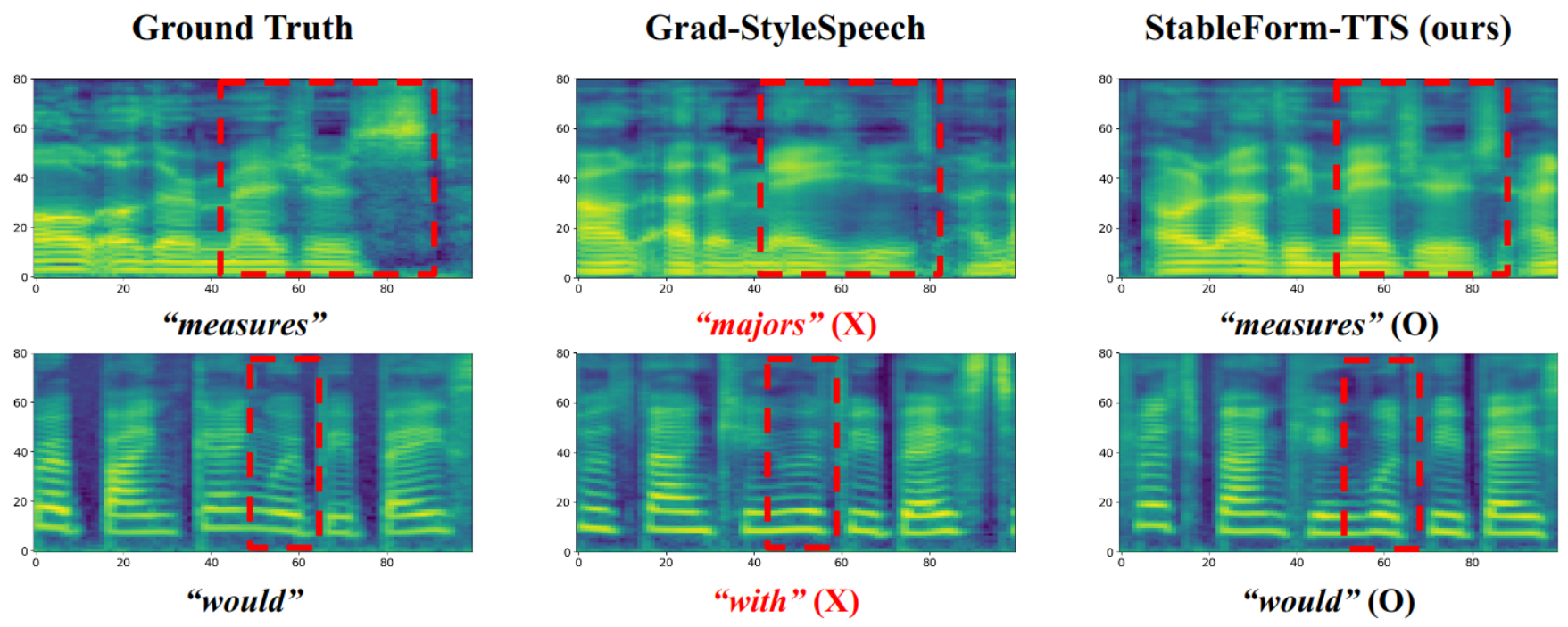}
    \caption{Visual comparison. The text beneath the mel-spectrogram indicates the ASR model's transcription of the area enclosed within the red dotted box.}
    \label{fig:visual_comparison}
\vspace{-0.5cm}
\end{figure}

\subsection{Experimental Results}
\subsubsection{Zero-Shot Speech Synthesis}
We compare the zero-shot speech synthesis performance of StableForm-TTS with other systems, which include i) Ground Truth, the ground truth audio; ii) Ground Truth (voc.), where we transform the ground truth audio into mel-spectrograms, then reconstruct the audio using the vocoder;  and iii) GSS, a cutting-edge zero-shot TTS model forming the foundation of our approach. 

As shown in Table \ref{tab:model_comparison}, StableForm-TTS outperforms GSS in terms of CER and WER, regardless of the training dataset. Notably, it surpasses the Ground Truth when trained on LT-460. Additionally, we plot the CER ratio in Fig. \ref{fig:preliminary_study}b to illustrate the relative increase in CER across different reverse steps, and it shows that our method demonstrates improved robustness compared to GSS. Our model also improves naturalness (UTMOS, WVMOS, MOSA-Net+, and MOS) and preserves speaker similarity (SECS and  SMOS) being on par with the  baseline. 

For visual inspection, we select two speakers from VCTK and plot the speech segments' mel-spectrograms of the ground truth, GSS, and StableForm-TTS. Fig. \ref{fig:visual_comparison} shows the pitch and formant contours of StableForm-TTS are clearly visible and similar to the ground truth. In contrast, GSS distorts the phonetic content, highlighted in the red dotted boxes, leading to incorrect pronunciations. 

\subsubsection{Ablation Study}
We perform ablation studies on our architecture using the two training datasets. As shown in Table \ref{tab:ablation_study}, the removal of either the E-F generators or the energy predictor results in salient performance drops. Therefore, splitting the acoustic modeling based on the source-filter theory and injecting not only the pitch but the energy, which was neglected in \cite{bak2021fastpitchformant}, are deemed beneficial. 

\subsubsection{Scalability Test}
We expand the English training datasets to 19,000 hours by including LT-R, LJSpeech\cite{ljspeech17}, DailyTalk\cite{lee2023dailytalk}, HiFi-TTS\cite{bakhturina2021hi}, Common Voice\cite{ardila2019common}, and MLS\cite{pratap2020mls}, and increase the model size by a factor of two (StableForm-large). Using the Coqui TTS toolkit\def\thefootnote{7}\footnote{\url{https://github.com/coqui-ai/TTS}}, we benchmark open-source zero-shot TTS models (Bark\def\thefootnote{8}\footnote{\url{https://github.com/suno-ai/bark}}, Tortoise\cite{betker2023better}, XTTS-v2\cite{casanova2024xtts}, and YourTTS\cite{casanova2022yourtts}), excluding VoiceCraft\cite{peng2024voicecraft}, which has an official implementation. Table \ref{tab:opensource_model_comparison} shows that StableForm-large either matches or outperforms leading zero-shot TTS benchmarks in terms of intelligibility, speaker similarity, and naturalness, while using substantially fewer parameters. Notably, the scale-up improves the pronunciation, and the effectiveness of our design is amplified compared to GSS-large. We expect applying large-scale vocoders will further improve the overall performance.

\begin{table}[t]
    \caption{Ablation Study}
    \label{tab:ablation_study}
    \resizebox{\columnwidth}{!}{
        \begin{tabular}{l|l|cccc|c|c}
        \toprule
        \textbf{Dataset} & \textbf{Model}  & \textbf{CER ($\downarrow$)} & \textbf{WER ($\downarrow$)} & \textbf{SECS ($\uparrow$)}  & \textbf{MOS-pred ($\uparrow$)} & \textbf{CMOS ($\uparrow$)}  & \textbf{\# Params} \\ \midrule
        \multirow{3}{*}{LT-460} 
        & StableForm-TTS          & \textbf{1.44 ± 0.05} & \textbf{3.64 ± 0.10} & 78.64 ± 0.07 & \textbf{4.249 ± 0.002} & 0.00 & 34.86M \\
        & w/o E-F generators      & 1.63 ± 0.05 & 4.05 ± 0.11 & \textbf{79.12 ± 0.06} & 4.235 ± 0.003 & -0.38 ± 0.05 & 34.78M \\ 
        & w/o Energy              & 1.87 ± 0.05 & 4.52 ± 0.11 & 77.87 ± 0.07 & 4.191 ± 0.003 & -0.39 ± 0.05 & 34.47M \\ \midrule
        \multirow{3}{*}{LT-R} 
        & StableForm-TTS          & \textbf{3.04 ± 0.08} & \textbf{6.74 ± 0.15} & \textbf{78.43 ± 0.07} & \textbf{4.060 ± 0.003} & 0.00 & 34.86M \\
        & w/o E-F generators      & 3.32 ± 0.08 & 7.35 ± 0.15 & 77.86 ± 0.07 & 4.057 ± 0.003 & -0.43 ± 0.05 & 34.78M \\ 
        & w/o Energy              & 4.09 ± 0.09 & 8.73 ± 0.18 & 77.01 ± 0.07 & 3.957 ± 0.003 & -0.39 ± 0.05 & 34.47M \\ \bottomrule 
        \end{tabular}
    }
\vspace{-0.4cm}
\end{table}

\begin{table}[t]
    \caption{Scalability Test. We use the PF solver with 10 reverse steps for StableForm-large. (Bold: best, Underline: second best)}
    \label{tab:opensource_model_comparison}
    \centering
    \resizebox{\linewidth}{!}{
        \begin{tabular}{l|cccc|c|c}
        \toprule
        \textbf{Model} & \textbf{CER ($\downarrow$)} & \textbf{WER ($\downarrow$)} & \textbf{SECS ($\uparrow$)} & \textbf{MOS-pred ($\uparrow$)} & \textbf{Hours} & \textbf{\# Params} \\
        \midrule
        Bark                & 5.38 ± 0.57 & 8.05 ± 0.69 & 74.64 ± 0.29 & 3.655 ± 0.011 & - & 1100.59M \\ 
        Tortoise            & \underline{0.44 ± 0.12} & \textbf{1.02 ± 0.16} & 76.51 ± 0.18 & \textbf{4.336 ± 0.006} & 49K & 972.57M \\
        VoiceCraft          & 1.72 ± 0.19 & 3.51 ± 0.29 & 79.05 ± 0.26 & 3.897 ± 0.007 & 20K & 839.61M \\
        XTTS-v2             & \textbf{0.43 ± 0.06} & \underline{1.51 ± 0.16} & \textbf{81.43 ± 0.16} & 4.219 ± 0.006 & 14K & 466.87M \\
        YourTTS             & 2.53 ± 0.17 & 5.83 ± 0.36 & 76.17 ± 0.20 & 4.031 ± 0.009 & 14K & 86.82M \\
        \midrule
        GSS-large & 1.80 ± 0.09 & 4.31 ± 0.20 & 77.93 ± 0.15 & 3.912 ± 0.003 & 19K & 68.13M \\
        StableForm-large    & 0.52 ± 0.08 & 1.55 ± 0.18 & \underline{79.26 ± 0.17} & \underline{4.333 ± 0.006} & 19K & 69.30M \\
        \bottomrule
        \end{tabular}
    }
\vspace{-0.5cm}
\end{table}

\section{Conclusion}
We propose StableForm-TTS, a diffusion-based zero-shot TTS system that surpasses the state-of-the-art baseline in terms of pronunciation accuracy and naturalness. Based on our preliminary findings, we effectively integrate the variance information and the source-filter theory into diffusion model to enhance pronunciation. StableForm-TTS minimizes the inherent mispronunciation issue in diffusion-based TTS by generating stable formant representation.


\begin{thebibliography}{00}

\bibitem{sohl2015deep} J. Sohl-Dickstein, E. Weiss, N. Maheswaranathan, and S. Ganguli, “Deep unsupervised learning using nonequilibrium thermodynamics,” in ICML, 2015.
\bibitem{ho2020denoising} J. Ho, A. Jain, and P. Abbeel, “Denoising diffusion probabilistic models,” in NeurIPS, 2020.
\bibitem{song2020score} Y. Song, J. Sohl-Dickstein, D. P. Kingma, A. Kumar, S. Ermon, and B. Poole, “Score-based generative modeling through stochastic differential equations,” in ICLR, 2020.
\bibitem{popov2021grad} V. Popov, I. Vovk, V. Gogoryan, T. Sadekova, and M. Kudinov, “Grad-tts: A diffusion probabilistic model for text-to-speech,” in ICML, 2021.
\bibitem{jeong2021diff} M. Jeong, H. Kim, S. J. Cheon, B. J. Choi, and N. S. Kim, “Diff-tts: A denoising diffusion model for text-to-speech,” arXiv preprint arXiv:2104.01409, 2021.
\bibitem{chen2020wavegrad} N. Chen, Y. Zhang, H. Zen, R. J. Weiss, M. Norouzi, and W. Chan, “Wavegrad: Estimating gradients for waveform generation,” in ICLR, 2020.
\bibitem{kong2020diffwave} Z. Kong, W. Ping, J. Huang, K. Zhao, and B. Catanzaro, “Diffwave: A versatile diffusion model for audio synthesis,” in ICLR, 2020.
\bibitem{chen2022resgrad} Z. Chen, Y. Wu, Y. Leng, J. Chen, H. Liu, X. Tan, Y. Cui, K. Wang, L. He, S. Zhao et al., “Resgrad: Residual denoising diffusion probabilistic models for text to speech,” arXiv preprint arXiv:2212.14518, 2022.
\bibitem{huang2022prodiff} R. Huang, Z. Zhao, H. Liu, J. Liu, C. Cui, and Y. Ren, “Prodiff: Progressive fast diffusion model for high-quality text-to-speech,” in ACM Multimedia, 2022.
\bibitem{chen2023lightgrad} J. Chen, X. Song, Z. Peng, B. Zhang, F. Pan, and Z. Wu, “Lightgrad: Lightweight diffusion probabilistic model for text-to-speech,” in ICASSP, 2023.
\bibitem{huang2022generspeech} R. Huang, Y. Ren, J. Liu, C. Cui, and Z. Zhao, “Generspeech: Towards style transfer for generalizable out-of-domain text-to-speech,” in NeurIPS, 2022.
\bibitem{casanova2022yourtts} E. Casanova, J. Weber, C. D. Shulby, A. C. Junior, E. G{\"o}lge, and M. A. Ponti, “Yourtts: Towards zero-shot multi-speaker tts and zero-shot voice conversion for everyone,” in ICML, 2022.
\bibitem{wang2023neural} C. Wang, S. Chen, Y. Wu, Z. Zhang, L. Zhou, S. Liu, Z. Chen, Y. Liu, H. Wang, J. Li et al., “Neural codec language models are zero-shot text to speech synthesizers,” arXiv preprint arXiv:2301.02111, 2023.
\bibitem{jiang2023mega} Z. Jiang, J. Liu, Y. Ren, J. He, Z. Ye, S. ji, Q. Yang, C. Zhang, P. Wei, C. Wang, X. Yin, Z. Ma et al., “Mega-tts 2: Boosting prompting mechanisms for zero-shot speech synthesis,” in ICLR, 2024.
\bibitem{chen2023diffusion} H. Chen and P. N. Garner, “Diffusion transformer for adaptive text-to-speech,” in SSW, 2023.
\bibitem{shen2023naturalspeech} K. Shen, Z. Ju, X. Tan, Y. Liu, Y. Leng, L. He, T. Qin, S. Zhao, and J. Bian, “Naturalspeech 2: Latent diffusion models are natural and zero-shot speech and singing synthesizers,” in ICLR, 2024.
\bibitem{kang2023grad} M. Kang, D. Min, and S. J. Hwang, “Grad-stylespeech: Any-speaker adaptive text-to-speech synthesis with diffusion models,” in ICASSP, 2023.
\bibitem{vaswani2017attention} A. Vaswani, N. Shazeer, N. Parmar, J. Uszkoreit, L. Jones, A. N. Gomez, Ł. Kaiser, and I. Polosukhin, “Attention is all you need,” in NeurIPS, 2017.
\bibitem{li2019neural} N. Li, S. Liu, Y. Liu, S. Zhao, and M. Liu, “Neural speech synthesis with transformer network,” in AAAI, 2019.
\bibitem{ren2020fastspeech} Y. Ren, C. Hu, X. Tan, T. Qin, S. Zhao, Z. Zhao, and T.-Y. Liu, “Fastspeech 2: Fast and high-quality end-to-end text to speech,” in ICLR, 2020.
\bibitem{badlani2022one} R. Badlani, A. Ła´ncucki, K. J. Shih, R. Valle, W. Ping, and B. Catanzaro, “One tts alignment to rule them all,” in ICASSP, 2022.
\bibitem{min2021meta} D. Min, D. B. Lee, E. Yang, and S. J. Hwang, “Meta-stylespeech: Multi-speaker adaptive text-to-speech generation,” in ICML, 2021.
\bibitem{chiba1941vowel} T. Chiba and M. Kajiyama, The vowel: Its nature and structure. Tokyo, Japan: Kaiseikan, 1941.
\bibitem{fant1960acoustic} G. Fant, Acoustic theory of speech production. The Hague, The Netherlands: Mouton, 1960.
\bibitem{tokuda2021source} I. Tokuda, “The source–filter theory of speech,” in Oxford Research Encyclopedia of Linguistics, 2021.
\bibitem{bak2021fastpitchformant} T. Bak, J. S. Bae, H. Bae, Y. I. Kim, and H. Y. Cho, “Fastpitchformant: Source-filter based decomposed modeling for speech synthesis,” in Interspeech, 2021.
\bibitem{ren2022revisiting} Y. Ren, X. Tan, T. Qin, Z. Zhao, and T.-Y. Liu, “Revisiting over-smoothness in text to speech,” in ACL, 2022.
\bibitem{lancucki2021fastpitch} A. {\L}a{\'n}cucki, “Fastpitch: Parallel text-to-speech with pitch prediction,” in ICASSP, 2021.
\bibitem{zen2019libritts} H. Zen, V. Dang, R. Clark, Y. Zhang, R. J. Weiss, Y. Jia, Z. Chen, and Y. Wu, “Libritts: A corpus derived from librispeech for text-to-speech,” in Interspeech, 2019.
\bibitem{koizumi2023libritts} Y. Koizumi, H. Zen, S. Karita, Y. Ding, K. Yatabe, N. Morioka, M. Bacchiani, Y. Zhang, W. Han, and A. Bapna, “Libritts-r: A restored multi-speaker text-to-speech corpus,” in Interspeech, 2023.
\bibitem{yamagishi2019vctk} J. Yamagishi, C. Veaux, and K. MacDonald, “Cstr vctk corpus: English multi-speaker corpus for cstr voice cloning toolkit (ver-sion 0.92),” 2019, https://doi.org/10.7488/ds/2645.
\bibitem{casanova2021sc} E. Casanova, C. Shulby, E. G{\"o}lge, N. M. M{\"u}ller, F. S. de Oliveira, A. C. Junior, A. d. S. Soares, S. M. Aluisio, and M. A. Ponti, “Sc-glowtts: an efficient zero-shot multi-speaker text-to-speech model,” in Interspeech, 2021.
\bibitem{Bernard2021} M. Bernard and H. Titeux, “Phonemizer: Text to phones transcription for multiple languages in python,” Journal of Open Source Software, vol. 6, no. 68, p. 3958, 2021. [Online]. Available: https://doi.org/10.21105/joss.03958
\bibitem{boersma1993accurate} P. Boersma, “Accurate short-term analysis of the fundamental frequency and the harmonics-to-noise ratio of a sampled sound,” in Proc. institute of phonetic sciences, 1993, pp. 97–110.
\bibitem{popov2021diffusion} V. Popov, I. Vovk, V. Gogoryan, T. Sadekova, M. S. Kudinov, and J. Wei, “Diffusion-based voice conversion with fast maximum likelihood sampling scheme,” in ICLR, 2021.
\bibitem{kong2020hifi} J. Kong, J. Kim, and J. Bae, “Hifi-gan: Generative adversarial networks for efficient and high fidelity speech synthesis,” in NeurIPS, 2020.
\bibitem{saeki2022utmos} T. Saeki, D. Xin, W. Nakata, T. Koriyama, S. Takamichi, and H. Saruwatari, “Utmos: Utokyo-sarulab system for voicemos challenge 2022,” in Interspeech, 2022.
\bibitem{andreev2023hifi} P. Andreev, A. Alanov, O. Ivanov, and D. Vetrov, “Hifi++: a unified framework for bandwidth extension and speech enhancement,” in ICASSP, 2023.
\bibitem{zezario2024mosa} R. E. Zezario, Y. W. Chen, S. W. Fu, Y. Tsao, H. M. Wang, and C. S. Fuh, "A study on incorporating whisper for robust speech assessment," in ICME, 2024.
\bibitem{lyth2024natural} D. Lyth and S. King, "Natural language guidance of high-fidelity text-to-speech with synthetic annotations," arXiv preprint arXiv:2402.01912, 2024.
\bibitem{ljspeech17} K. Ito and L. Johnson, “The lj speech dataset,” https://keithito.com/LJ-Speech-Dataset/, 2017.
\bibitem{lee2023dailytalk} K. Lee, K. Park, and D. Kim, “Dailytalk: Spoken dialogue dataset for conversational text-to-speech,” in ICASSP, 2023.
\bibitem{bakhturina2021hi} E. Bakhturina, V. Lavrukhin, B. Ginsburg, and Y. Zhang, “Hi-fi multi-speaker english tts dataset,” arXiv preprint arXiv:2104.01497, 2021.
\bibitem{ardila2019common} R. Ardila, M. Branson, K. Davis, M. Kohler, J. Meyer, M. Henretty, R. Morais, L. Saunders, F. Tyers, and G. Weber, “Common voice: A massively-multilingual speech corpus,” in LREC, 2020.
\bibitem{pratap2020mls} V. Pratap, Q. Xu, A. Sriram, G. Synnaeve, and R. Collobert, “Mls: A large-scale multilingual dataset for speech research,” in Interspeech, 2020.
\bibitem{betker2023better} J. Betker, “Better speech synthesis through scaling,” arXiv preprint arXiv:2305.07243, 2023.
\bibitem{casanova2024xtts} E. Casanova, K. Davis, E. G{\"o}lge, G. G{\"o}knar, I. Gulea, L. Hart, A. Aljafari, J. Meyer, R. Morais, S. Olayemi et al., “Xtts: a massively multilingual zero-shot text-to-speech model,” in Interspeech, 2024.
\bibitem{peng2024voicecraft} P. Peng, P.-Y. Huang, D. Li, A. Mohamed, and D. Harwath, “Voicecraft: Zero-shot speech editing and text-to-speech in the wild,” in ACL, 2024.

\end{thebibliography}
\end{document}